\RequirePackage{etoolbox}
\csdef{input@path}{%
 {sty/}
 {img/}
}%
\csgdef{bibdir}{bib/}

\documentclass{imsart}
\usepackage{amsthm}
\usepackage{bm}
\usepackage{amsmath}
\usepackage{natbib}
\usepackage[colorlinks,citecolor=blue,urlcolor=blue,filecolor=blue,backref=page]{hyperref}
\usepackage{graphicx}

\newcommand{\bsP}{\boldsymbol{\Psi}}
\newcommand{\mbN}{\mathbf{N}}

\newcommand{\mbU}{\mathbf{U}}
\newcommand{\mbx}{\mathbf{x}}
\newcommand{\mbL}{\mathbf{\Lambda}}

\startlocaldefs
\numberwithin{equation}{section}
\theoremstyle{plain}

\endlocaldefs

\begin{document}

\begin{frontmatter}
\title{Bayesian Combinatorial Multi-Study \\ Factor Analysis}
\runtitle{Bayesian Combinatorial MSFA}

\begin{aug}
\author{\fnms{Isabella N.} \snm{Grabski}\thanksref{addr1}},
\author{\fnms{Roberta De} \snm{Vito}\thanksref{addr2}},
\author{\fnms{Lorenzo} \snm{Trippa}\thanksref{addr1,addr3}}

\and
\author{\fnms{Giovanni} \snm{Parmigiani}\thanksref{addr1,addr3}}

\runauthor{I. N. Grabski et al.}

\address[addr1]{Department of Biostatistics,
     Harvard University,
     Boston, MA 
}

\address[addr2]{Department of Biostatistics and Data Science Initiative, Brown University, Providence, RI
}

\address[addr3]{Department of Data Sciences,
Dana-Farber Cancer Institute,
Boston, MA
}

\end{aug}

\begin{abstract}
Analyzing multiple studies allows leveraging data from a range of sources and populations, but until recently, there have been limited methodologies to approach the joint unsupervised analysis of multiple high-dimensional studies. A recent method, Bayesian Multi-Study Factor Analysis (BMSFA), identifies latent factors common to all studies, as well as latent factors specific to individual studies. However, BMSFA does not allow for partially shared factors, i.e. latent factors shared by more than one but less than all studies. We extend BMSFA by introducing a new method, Tetris, for Bayesian combinatorial multi-study factor analysis, which identifies latent factors that can be shared by any combination of studies. We model the subsets of studies that share latent factors with an Indian Buffet Process. We test our method with an extensive range of simulations, and showcase its utility not only in dimension reduction but also in covariance estimation. Finally, we apply Tetris to high-dimensional gene expression datasets to identify patterns in breast cancer gene expression, both within and across known classes defined by germline mutations.
\end{abstract}



\end{frontmatter}


    

    


\section{Introduction}
Using multiple studies in a single statistical analysis is a powerful way to leverage data from a variety of sources and populations. This is particularly relevant with high-throughput biological data, where analyses are often carried out using systematic collections of genomic data. These data are generated over time in different laboratories, with different techniques and/or under different experimental conditions, and it has been well-established that such data contain both biological and technological sources of variation \citep{irizarry2003exploration}. Methods that jointly analyze multiple studies offer the opportunity to distinguish biological signals from artifacts by identifying what signal is shared by some or all of the studies, and what signal is specific to an individual study. These types of analyses and the resulting challenges are not unique to high-throughput biology. Methods producing replicable features and patterns across heterogeneous settings are valuable to many disciplines.

Many of the datasets studied in these contexts, especially in high-throughput biology, are high-dimensional, which creates an extra challenge for their joint analysis. The unsupervised identification of latent factors can be particularly useful for uncovering signal in the high-dimensional setting, but there has been limited work on formal statistical approaches to unsupervised learning for multiple studies. De~Vito {\it et al.} presented the first extension of factor analysis to the multi-study context \citep{de2019multi}, and more recently developed a Bayesian multi-study factor analysis methodology (BMSFA) that is better suited for high-dimensional data \citep{de2018bayesian}. 

However, these approaches can only identify latent factors if they are common to all studies or unique to a single study. In several applications, it is important to also identify biological factors only shared by subsets of studies. For example, in genomics, we might expect the subset of studies using the same experimental techniques and/or generated from the same laboratory to share common latent factors. 

In this work, we introduce a Bayesian combinatorial multi-study factor analysis method, which extends the Bayesian multi-study factor analysis method \citep{de2018bayesian} to learn latent factors shared by any subset of studies. We do so by using the Indian Buffet Process \citep{ghahramani2006infinite} to model the shared ownership of factors across multiple studies, which encourages sparsity. We thus refer to our method as Tetris, since the matrix indicating factor ownership resembles patterns from the Tetris video game. We estimate parameters using a Gibbs sampling algorithm tuned to ensure feasibility in $p\gg n$ settings. We test Tetris through a broad range of simulations, and we highlight Tetris's utility not only in dimension reduction but also in covariance estimation. 

While we present our methods and simulations in the framework of multi-study analysis, our model is equally relevant for a single study which examines multiple subgroups or conditions that may or may not share a common structure. We will illustrate this case in our breast cancer application, where we apply Tetris to four high-dimensional RNA-sequencing datasets to identify patterns in lympoblastoid cell gene expression, both within and across genotypes related to breast cancer risk.    

\section{Methods}

\subsection{Model and Estimation}

We consider $S$ studies with the same $P$ variables. Each study $s$ has $n_s$ subjects and $P$-dimensional centered data vector $\mbx_{is}$ with $i =1, \dots, n_s$. If $K$ is the total number of factors (including common, study-specific, and partially shared), then our model is \begin{equation} \mbx_{is} = \mbL \bm{A}_s \bm{l}_{is} + \bm{e}_{is} \end{equation} where $\bm{\Lambda}$ is a $P \times K$ factor loadings matrix and $\bm{l}_{is}$ are $K \times 1$ latent factors.

The $K \times K$ study-level factor indicator matrix $\bm{A}_s$ is our key element to estimate the pattern of partial sharing of factors. This matrix consists of all 0s, except for the diagonal entries, which are either 1 or 0. The $k$th diagonal entry of $\bm{A}_s$ is 1 whenever the $k$th factor is present in study $s$, so the product $\bm{\Lambda} \bm{A}_s \bm{l}_{is}$ will include the corresponding elements. Hence, $\bm{A}_s$ controls which factors are included and which are not into the model for study $s$. 

We denote by $\mathcal{A}$ the overall $S \times K$ factor indicator matrix whose $s$th row consists of the $K$ diagonal entries of $\bm{A}_s$. If the $k$th column of $\mathcal{A}$ consists of all 1s, the $k$th factor is a common factor; if it consists of exactly a single 1, this is a study-specific factor; and if it consists of at least one 1 and one 0, this is a partially shared factor. In this model, we enforce a minimum of one study-specific factor per study by having each of the first $S$ columns correspond to a study-specific factor for each. In other words, the first $S$ columns of $\mathcal{A}$ form the $S \times S$ identity matrix. This decision is motivated by considering each study to have some batch effect that it uniquely expresses, and allows the other factors discovered to be interpreted conditional on separating out these batch effects.

Our prior model builds on BMSFA \citep{de2018bayesian}. Namely, the latent factors are $\bm{l}_{is} \sim \mathcal{N}_{K}(0,\bm{I}_k)$, and the error terms are $\bm{e}_{is} \sim \mathcal{N}_{P}(0, \bsP_s)$ for $\bsP_s = diag(\psi_{s1}^2, \ldots, \psi_{sP}^2).$ In turn, the elements of this matrix are $\psi_{sp}^{-2} \sim \Gamma(a_{\psi}, b_{\psi})$. For the loading matrix $\bm{\Lambda}$ we use the multiplicative gamma process shrinkage prior \citep{bhattacharya2011sparse}, to encourage the definition of factors with decreasing norm, i.e., for each loading matrix element $\Lambda_{pk} \sim \mathcal{N}(0, \omega_{pk}^{-1} \tau_{k}^{-1})$ with $\tau_k = \prod_{l=1}^k \delta_l$,  $\omega_{pk} \sim \Gamma\left(\frac{\nu}{2},\frac{\nu}{2}\right)$, $\delta_1 \sim \Gamma(a_1,1)$, and $\delta_l \sim \Gamma(a_2,1)$ for $l \geq 2$.

The study-level factor indicator matrices $\bm{A}_s$, which are jointly summarized by the overall factor indicator matrix $\mathcal{A}$, allow us to estimate factors shared by any subset of studies. To avoid having to select the number of factors ahead of time, and to avoid the explicit assessment of all combinatorial possibilities, we place an Indian Buffet Process prior on $\mathcal{A}.$ To be precise, since the first $S$ columns are fixed as the $S \times S$ identity matrix as described above, the Indian Buffet Process (IBP) prior is placed on $\mathcal{A}$ beginning from the $S+1$th column. 

The IBP \citep{ghahramani2006infinite} is a probability distribution defined over infinite binary matrices, specifically matrices that have a finite number of rows and an infinite number of columns. In our case, each row represents a study, and each column represents a factor; as described earlier, an entry of 1 indicates that a given factor is shared by the corresponding study. Although the number of columns is infinite, the expected number of columns whose entries are not all 0 is finite.  This expectation increases with the number of rows (i.e., studies). Hence, using this prior on $\mathcal{A}$ implies that we will automatically be performing dimension selection, without having to use heuristic post-hoc measures to determine the number of factors. The resulting patterns of 1s and 0s in this matrix resemble the patterns of fallen Tetris blocks, giving our method its name.

We specifically choose to use the two-parameter generalization of the IBP \citep{knowles2007infinite}. In the regular one-parameter model, a single parameter $\alpha$ governs both the expected number of nonzero columns and the expected sparsity of each column. However, in the two-parameter version, $\alpha$ governs the expected sparsity of the total matrix, whereas $\alpha$ and $\beta$ together govern the expected number of columns. The resulting effect is that for fixed $\alpha$, small values of $\beta$ are more likely to result in factors shared by larger numbers of studies (i.e., more common factors, and factors shared by large subsets), whereas large values of $\beta$ are more likely to result in factors shared by small numbers of studies (i.e., more study-specific factors, and factors shared by small subsets). This is a desirable property in our setting because, depending on the application, data may fall at one end of the spectrum or the other. 

To sample from the posterior distributions using these priors and our model, we present the following computationally efficient Gibbs sampler (based primarily on the work in \cite{de2018bayesian}, \cite{knowles2007infinite}, and \cite{doshi2009indian}):

\begin{enumerate}
\item We follow \cite{knowles2007infinite} to update each entry of $\mathcal{A}$. For $s$ ranging from 1 to $S$, for $k$ ranging from $S+1$ to $K$, set $\mathcal{A}_{sk} = 1$ with probability $r / (r+1) $, where 
\begin{multline*} 
r = \tilde{m} \times \prod_{i=1}^{n_s} \exp\left\{ -\frac{1}{2} \left( (\bm{x}_{is} - \bm{\Lambda}\bm{A}_s^{k1}\bm{l}_{is})^T \Psi_{s}^{-1} (\bm{x}_{is} - \bm{\Lambda}\bm{A}_s^{k1}\bm{l}_{is}) \; -
\right. \right.\\ 
\left. (\bm{x}_{is} - \bm{\Lambda}\bm{A}_s^{k0} \bm{l}_{is})^T \Psi_s^{-1} (\bm{x}_{is} - \bm{\Lambda}\bm{A}_s^{k0} \bm{l}_{is}) )\right\}, 
\end{multline*} 
and $\tilde{m} = \mbox{Pr}(\mathcal{A}_{sk}=1\mid \cdots) / \mbox{Pr}(\mathcal{A}_{sk}=0 \mid \cdots) = m_{k,-s}/(\beta + S - 1 - m_{k,-s})$. Here, $m_{k,-s}$ denotes the sum of column $k$ of $\mathcal{A}$, subtracting the current value of $\mathcal{A}_{sk}$. Further, $\bm{A}_s^{k1}$ has the $k$th diagonal element of $\bm{A}_s$ equal to 1 (i.e. $\mathcal{A}_{sk}=1$) and $\bm{A}_s^{k0}$ has the $k$th diagonal element of $\bm{A}_s$ equal to 0 (i.e. $\mathcal{A}_{sk}=0$)
\item Again as in \cite{knowles2007infinite} and as discussed by \cite{doshi2009indian}, we use a Metropolis-Hastings step to allow new factors to be added. For $s$ ranging from 1 to $S$, to sample the number of new factors $k_{\text{new}}$ for study $s$, sample $k_{\text{new}} \sim \text{Pois}\left(\frac{\alpha \beta}{\beta + S - 1}\right)$, and the corresponding new elements $\omega_{\text{new}}$, $\delta_{\text{new}}$, and $\bm{l}_{s, \text{new}}$ from their respective priors outlined in the previous section. Then accept all these samples with probability $\min(1,r)$, for $r$ equal to 
\begin{align*} \prod_{p=1}^P &|2\pi \bm{D}_{p, \text{new}}|^{-\frac{1}{2}} |2\pi (\bm{D}_{p,\text{new}}^{-1} + \psi_{sp}^{-2} \bm{A}_{s,\text{new}} \bm{l}^T_{s,\text{new}} \bm{l}_{s,\text{new}} \bm{A}_{s,\text{new}})^{-1}|^{\frac{1}{2}} \\ &\times \exp\left\{\frac{1}{2}\bar{\bm{\Lambda}}_{p,\text{new}} (\bm{D}_{p,\text{new}}^{-1} + \psi_{sp}^{-2} \bm{A}_{s,\text{new}} \bm{l}^T_{s,\text{new}} \bm{l}_{s,\text{new}} \bm{A}_{s,\text{new}})\bar{\bm{\Lambda}}^T_{p,\text{new}} \right \}. \end{align*}
This is equivalent to the ratio of the likelihoods with and without the proposed factors, with the new elements of $\Lambda$ marginalized out. Here, we define $\bar{\bm{\Lambda}}^T_{p,\text{new}}$ as  $$(\bm{D}_{p,\text{new}}^{-1} + \psi_{sp}^{-2}\bm{A}_{s,\text{new}} \bm{l}^T_{s,\text{new}} \bm{l}_{s,\text{new}} \bm{A}_{s,\text{new}})^{-1} \psi_{sp}^{-2} \bm{A}_{s,\text{new}} \bm{l}^T_{s,\text{new}}(\bm{x}_s^{(p)} + \bm{l}_s \bm{A}_s \bm{\Lambda}^T_p).$$
\item To update the loadings matrix elements and the associated parameters in the next 4 steps, we follow the main ideas of \cite{bhattacharya2011sparse} and the multi-study extension from \cite{de2018bayesian}. For $p$ ranging from 1 to $P$, sample the transpose of the $p$th row of the factor loadings matrix 
\begin{align*} \begin{split}
\bm{\Lambda}_p^T \sim \mathcal{N}\{\left(\bm{D}_p^{-1} + \sum_{s=1}^S \psi_{sp}^{-2} \bm{A}_s \bm{l}_s^T \bm{l}_s \bm{A}_s \right)^{-1} &\left(\sum_{s=1}^S \psi_{sp}^{-2} \bm{A}_s \bm{l}_s^T \bm{x}_s^{(p)}\right),\\ &  \left(\bm{D}_p^{-1} + \sum_{s=1}^S \psi_{sp}^{-2} \bm{A}_s \bm{l}_s^T \bm{l}_s \bm{A}_s \right)^{-1}  \},\end{split} \end{align*} where $\bm{l}_s$ is the $n_s \times K$ matrix of latent factors across samples for study $s$, $\bm{x}_s^{(p)}$ is the $n_s \times 1$ vector of values for feature $p$ across samples for study $s$, and $\bm{D}_p^{-1} = \text{diag}(\omega_{p1}\tau_1, \ldots, \omega_{pK}\tau_K)$.
\item For $s$ ranging from 1 to $S$, $i$ ranging from 1 to $n_s$, sample \begin{align*} \begin{split} \bm{l}_{is} \sim \mathcal{N}\{(\bm{I}_K + \bm{A}_s \bm{\Lambda}^T \bsP_s^{-1} \bm{\Lambda} \bm{A}_s)^{-1}(\bm{A}_s &\bm{\Lambda}^T \Psi_s^{-1}\bm{x}_{is}), \\ & (\bm{I}_K + \bm{A}_s \bm{\Lambda}^T \bsP_s^{-1} \bm{\Lambda} \bm{A}_s)^{-1}\}.\end{split} \end{align*} 
This samples a $K \times 1$ vector regardless of how many factors are shared by study $s$. If the $k$th factor is not shared by the study, then the $k$th diagonal element of $\bm{A}_s$ will be 0; as a result, this sampling step effectively draws the corresponding element of $\bm{l}_{is}$ from the prior, and that element of $\bm{l}_{is}$ will not affect the likelihood due to multiplication by $\bm{A}_s$.
\item For $p$ ranging from 1 to $P$, for $k$ ranging from 1 to $K$, sample 
\[
\omega_{pk} \sim \Gamma\left(\frac{\nu+1}{2}, \frac{\nu + \tau_k \Lambda_{pk}^2}{2}\right).
\]
\item Sample $\delta_1 \sim \Gamma\left(a_1 + \frac{PK}{2}, 1 + \frac{1}{2} \sum_{k=1}^K \tau_k^{(1)} \sum_{p=1}^P \omega_{pk} \Lambda_{pk}^2\right)$, where $\tau_{k}^{(1)}$ represents $\dfrac{\tau_k}{\delta_1}$.
\item For $l$ ranging from 2 to $K$, sample 
\[
\delta_l \sim \Gamma\left(a_2 + \frac{P}{2}(K-l+1), 1 + \frac{1}{2} \sum_{k=l}^K \tau_k^{(l)} \sum_{p=1}^P \omega_{pk} \Lambda_{pk}^2\right),\] where $\tau_{k}^{(l)}$
represents $\dfrac{\tau_k}{\delta_l}$. 
\item For $s$ ranging from 1 to $S$, for $p$ ranging from 1 to $P$, sample \begin{equation*} \psi_{sp}^{-2} \sim \Gamma\left(a_{\psi} + \frac{n_s}{2}, b_{\psi} + \frac{1}{2} \sum_{i=1}^{n_s} (\bm{x}_{is}^{(p)} - \bm{\Lambda}_p \bm{A}_{s} \bm{l}_{is})^2\right). \end{equation*}
\end{enumerate}

\subsection{Recovery of Factor Indicator Matrix and Loading Matrix}

In the sampler described in the previous section, the number of factors can change dynamically from iteration to iteration in the chain. This means that the MCMC samples of the factor indicator matrix $\mathcal{A}$ and of the loading matrix $\Lambda$ can have varying dimensions across iterations. Recovering point estimates of these quantities requires post-processing on the sampler output.

To recover $\mathcal{A}$, we recommend the following approach, which we use in all results presented in this work. First, we define a distance between two sampled matrices $\mathcal{A}_i, \mathcal{A}_j$ as the minimum number of $0 \to 1$ or $1 \to 0$ ``flips" needed to change all the entries of $\mathcal{A}_i$ to those of $\mathcal{A}_j$ (or, symmetrically, from $\mathcal{A}_j$ to $\mathcal{A}_i$) over all possible permutations of their columns. Intuitively, this counts the number of differences between the two matrices under the best possible ``alignment" of their columns, since the ordering of the columns (which correspond to factors) is not itself meaningful. We can express this formally as $$\min_{\boldsymbol{L,R}} \text{tr}(\boldsymbol{LMR}),$$ where each $M_{kl}$ is the number of flips between the $k$th column of $\mathcal{A}_i$ and the $l$th column of $\mathcal{A}_j$, and $\boldsymbol{L}, \boldsymbol{R}$ represent left and right permutation matrices respectively. This can be efficiently computed using the Hungarian Algorithm, as implemented in the library \texttt{clue} \citep{hornik2005clue}. 

We compute these distances between every pair of post-burn-in MCMC samples, and then select the MCMC sample that contains the largest number of other MCMC samples within a radius $r$. We choose $r$ as whichever is larger between the 0.05th quantile of all distances and the number of studies $S$. The motivation for this upper bound on the radius is to allow each study to change one entry, but other options could be used instead. Overall, we can think of this selection process as choosing the sampled $\mathcal{A}$ matrix that defines the highest density neighborhood when chosen as the center. Ties are broken first by selecting the matrix with fewest factors, and if there are still ties, then by selecting the matrix with highest probability under the IBP prior.  

To summarize the uncertainty around this point estimate of $\mathcal{A}$, we adapt the idea of credible balls from the Bayesian clustering context \citep{wade2018bayesian}. Following their work, we define a credible ball of radius $\epsilon$ as $$B_{\epsilon}(\mathcal{A}^{*}) = \{ \mathcal{A}: d(\mathcal{A},\mathcal{A}^{*})\leq \epsilon \}$$ for point estimate $\mathcal{A}^{*}$. Then we define a level $1-\alpha$ credible ball as $B_{\epsilon^{*}}(\mathcal{A}^{*})$ with $\epsilon^{*}$ the smallest $\epsilon \geq 0$ such that $$\mbox{Pr}(B_{\epsilon}(\mathcal{A}^{*})|\boldsymbol{X}) \geq 1-\alpha$$ for data $\boldsymbol{X}$. In practice, we can approximate this quantity using the MCMC output as $\frac{1}{M}\sum_{m=1}^M \boldsymbol{1}(d(\mathcal{A}^m,\mathcal{A}^{*}) \leq \epsilon),$ if there are $M$ iterations post-burn-in and $\mathcal{A}^m$ represents the $m$-th such sampled matrix. 

Next, to recover $\bm{\Lambda}$, we first need sampler output where $\mathcal{A}$ is exactly the same across all iterations; otherwise, it is not meaningful to construct a single estimate of $\bm{\Lambda}$, since there would not be a single set of factors for the columns of $\bm{\Lambda}$ to correspond to. Hence, after recovering the point estimate $\mathcal{A}$, we propose to rerun the MCMC with $\mathcal{A}$ fixed to this estimate, while updating all the remaining parameters as previously. 

In factor analysis, loading matrices are not unique since they are only identifiable up to rotations. Different draws of $\bm{\Lambda}$ from iteration to iteration could correspond to a different rotation of the same parameter. Thus, to estimate $\bm{\Lambda}$, we cannot simply take, say, the posterior median of all $\bm{\Lambda}$ draws in this new chain. 
We address this issue using a strategy that extends that of BMSFA \citep{de2018bayesian}. We first partition factors into ``types" depending on their sharing patterns in the estimate of $\mathcal{A}$. For example, common factors would be one type. Factors shared by studies 3 and 5 would be another. We then compute the median loading matrix covariance corresponding to each type of factor. For example, the common loading matrix covariance is computed as the covariance when $\Lambda$ is subsetted to only include columns corresponding to the common factors, and similarly for all other types. Although there are potentially many types of factors, their number growing exponentially in the number of studies, in practice the estimation of $\mathcal{A}$ is parsimonious, so the number of unique types of factors to be considered is generally computationally feasible. For each resulting loading matrix covariance $\hat{\tilde{\bm{\Lambda}}}\hat{\tilde{\bm{\Lambda}}}^\top$, where $\tilde{\bm{\Lambda}}$ represents the loading matrix subsetted to the appropriate columns, we use the matrix spectral decomposition to decompose \begin{equation*} \hat{\tilde{\bm{\Lambda}}}\hat{\tilde{\bm{\Lambda}}}^\top = \mbU \mbN \mbU^\top = (\mbU \mbN^{\frac{1}{2}})(\mbN^{\frac{1}{2}}\mbU)^\top, \end{equation*} and we then take our estimate of $\tilde{\bm{\Lambda}}$ as $\hat{\tilde{\bm{\Lambda}}} = \mbU \mbN^{\frac{1}{2}}.$ All columns of $\bm{\Lambda}$ can be estimated in this way. 

\section{Simulations}

\subsection{Simulation Design}

We evaluated our method in three different simulated scenarios, with a range of parameters encompassing similar dimensionality, sample size, and number of studies as our real data application. In brief, the first scenario specifically assesses Tetris's ability to differentiate signal that is partially shared from the signal that is common to all studies. The second scenario is designed to test Tetris's performance as the data dimensions, loading matrix sparsity, and number of partially shared factors are systematically varied. Finally, the third scenario evaluates Tetris's performance when the number of studies is greatly increased. These scenarios are loosely based on those used to evaluate BMSFA \citep{de2018bayesian}, and modified to study partially shared factors. We use ``partially shared" to refer to factors belonging to multiple, but not all, studies. In all of the following scenarios, each specified combination of settings was simulated ten times, and Tetris was run with a total of 10,000 iterations and a burn-in of 8,000 iterations. These iterations were sufficient for convergence and good mixing.

\subsubsection{Scenario 1: Structurally Distinct Common and Partially Shared Signals}

In the first scenario, we simulated four studies $\bm{X}_s$ with $n$ observations and $p$ variables from $\mathcal{MVN}(\bm{0},\Sigma_s)$, with $\Sigma_s = \Lambda \bm{A}_s \Lambda^T + \Psi_s.$ We fixed three common factors and three factors shared by the first two studies, as well as one study-specific factor per study. 

The non-zero elements of $\Lambda$ were drawn from $\mathcal{U}(-1,1)$, but for the columns corresponding to the common factors, their locations were chosen uniformly at random among the first $\frac{p}{2}$ rows (i.e. these loadings only involved the first half of the variables), and for the columns corresponding to the partially shared factors, their locations were chosen uniformly at random among the last $\frac{p}{2}$ rows (i.e. these loadings only involved the second half of the variables). This was designed to create structural distinction between the common and partially shared factors. The locations of the non-zero elements for columns corresponding to study-specific factors were not restricted, and were selected uniformly at random over all rows. The number of such locations varied with the sparsity of the non-structurally-zero elements of the loading matrix, which was set to either $80\%$ or $50\%$. 

$\Psi_s$ is a diagonal matrix where each element is drawn from $\mathcal{U}(0,0.5).$ We set the dimension of the data $(n,p)$ to our $p \gg  n$ setting $(10,60)$.  

\subsubsection{Scenario 2: Inference on Dimension, Sparsity, and Number of Partially Shared Factors}

In the second scenario, we simulated four studies $\bm{X}_s$ with $n$ samples and $p$ variables each by drawing from $\mathcal{MVN}(\bm{0},\Sigma_s)$, where $\Sigma_s = \Lambda \bm{A}_s \Lambda^T + \Psi_s.$ Each non-zero element of $\Lambda$ was drawn from $\mathcal{U}(-1,1)$, and their locations in the matrix were selected uniformly at random. $\Psi_s$ is a diagonal matrix where each element is drawn from $\mathcal{U}(0,0.5).$ All simulations in this scenario have three common factors and one study-specific factor per study. 

We then varied three parameters: the dimension of the data, where $(n,p)$ is set to one of $(60,10), (35,35)$, and $(10,60)$ to correspond to $p\ll n$, $p=n$, and $p\gg n$ settings; the sparsity of the loading matrix, which is one of $20\%, 50\%,$ and $80\%$; and the number of partially shared factors, which is either zero, one (shared by the first two studies), or two (both shared by the first two studies). In total, we considered all 27 combinations of these parameters under this scenario.

\subsubsection{Scenario 3: Large Number of Studies}

In the third scenario, we adapted the second scenario to consider the case where the data consists of 16 studies. We used the $p\gg n$ (i.e. $(n,p) = (10,60)$) and 80\% sparsity setting, and varied the number of partially shared factors as 0 or 1. When there is one partially shared factor, that factor is shared by the first eight studies. 

\subsection{Evaluation Metrics}

We used two primary metrics to evaluate the accuracy of our simulation results. 

\subsubsection{RV Coefficient of Full Loading Matrix Covariance}

We quantified the similarity between the true and estimated full loading matrix covariance $\Lambda \Lambda^T$ using the RV coefficient \citep{abdi2007rv}. For two positive semi-definite matrices $S$ and $T$, the RV is defined as 
\begin{equation} RV(S,T) = \frac{\mathrm{Tr}\hspace{1pt}(S^T T)}{\sqrt{\mathrm{Tr}\hspace{1pt}(S^T S)\mathrm{Tr}\hspace{1pt}(T^T T)}}, \end{equation}
and takes on a value between 0 and 1, where a value closer to 1 indicates greater similarity between the two matrices. Note that the quantity $\Lambda \Lambda^T$ can be interpreted as summarizing relationships among variables for a hypothetical study containing all the factors found across any of the observed studies. 

\subsubsection{Reconstruction of Study Covariances}

We reconstructed the study-specific covariance matrix of study $s$ as $\hat{\Sigma}_{s,Tetris} = \hat{\Lambda} \bm{\hat{A}}_s \hat{\Lambda}^T$. We then computed the RV coefficient between this matrix and the true covariance matrix used to generate the data. As a comparison, we applied the same procedure to estimates obtained from BMSFA via the \texttt{MSFA} package \citep{de2018bayesian}, as well as standard, single-study factor analysis (FA) via the \texttt{psych} package \citep{revelle2017psych}. 

BMSFA separately estimates a common loading matrix covariance and a study-specific loading matrix for each study. When reconstructing the covariance matrix with BMSFA, we evaluated the sum of this estimated common loading matrix covariance and study-specific loading matrix covariance. We obtained these estimates as the posterior medians of the common and study-specific loading matrix covariances from each post-burn-in iteration of the MCMC chain, subsetting the appropriate loading matrix in each iteration to BMSFA's chosen number of factors. 

When using FA, which only supports the analysis of one study at a time, we considered each study independently and estimated $\hat{\Sigma}_{s,FA} = \hat{\Lambda} \hat{\Lambda}^T$, that is the estimated loading matrix covariance. Since this method requires specifying the number of factors, we used the ground truth number of factors, thus considering the least favorable scenario for our method. Neither Tetris nor BMSFA  require specifying the number of factors.

\subsection{Simulation Results}

We first examine the results of the Scenario 1 simulation to highlight Tetris's ability to clearly and consistently discriminate between common and partially shared signal, when there are important structural differences between the two. We use ``common" to refer to signal that is shared by all the studies under consideration, and ``partially shared" to refer to signal that is shared by more than one but not all of the studies. In this particular simulation, the partially shared factors are all shared by the first two (out of four) studies. The structural differences between the common and partially shared factors are that the set of features involved in the common factors and the set of features involved in the partially shared factors are disjoint and exhaustive. Note that the study-specific factors for each of the four studies may span the entire set of features.

\begin{figure}[t]
\begin{center}
\begin{tabular}{ c c c }
& \textbf{Ground Truth} & \textbf{Estimated} \\
 \textbf{(A)} & & \\
& \includegraphics[width=0.43\textwidth]{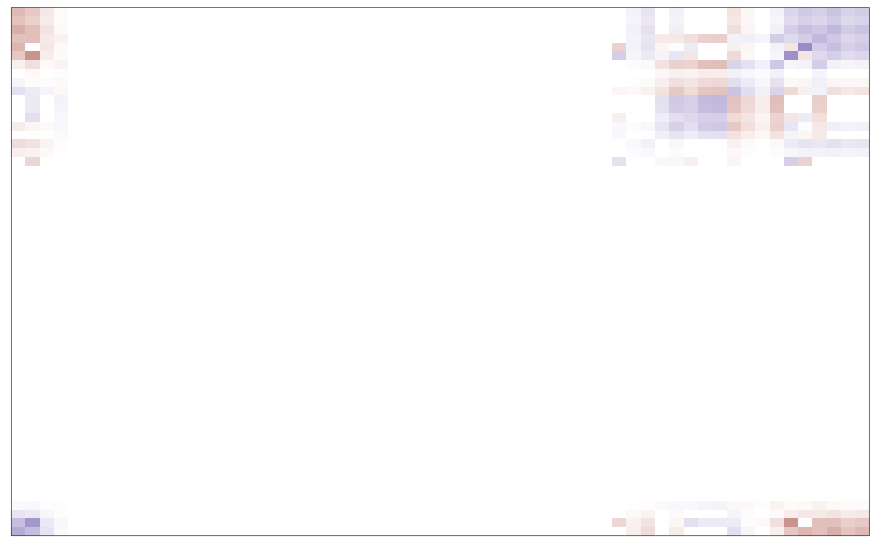} & \includegraphics[width=0.43\textwidth]{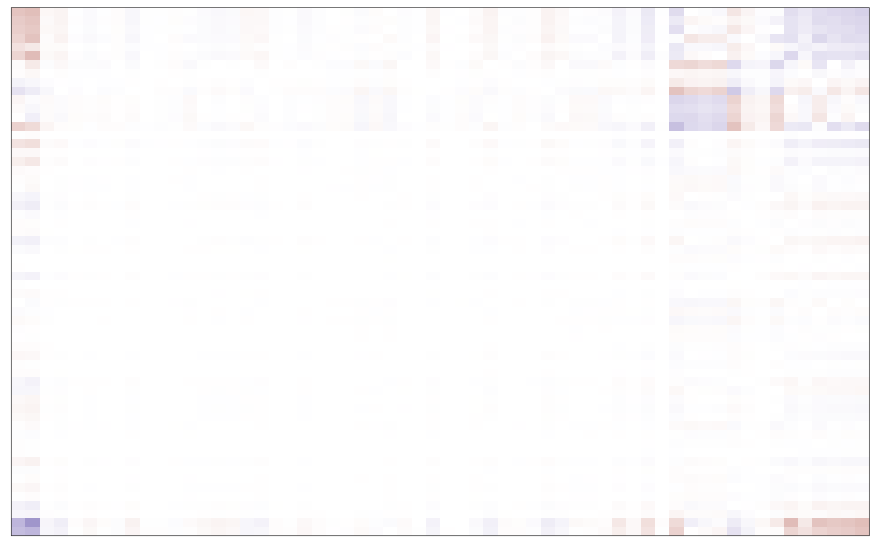}  \\ 
  \textbf{(B)}&& \\
 & \includegraphics[width=0.43\textwidth]{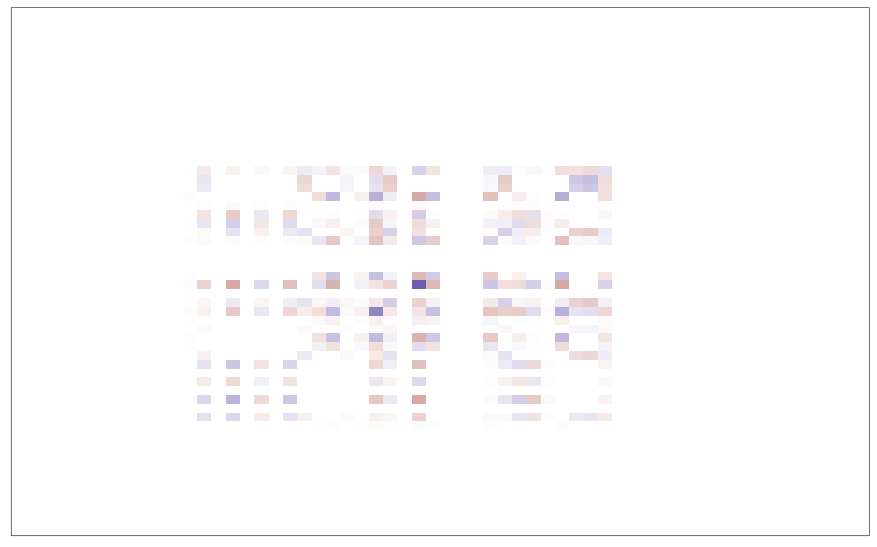} & \includegraphics[width=0.43\textwidth]{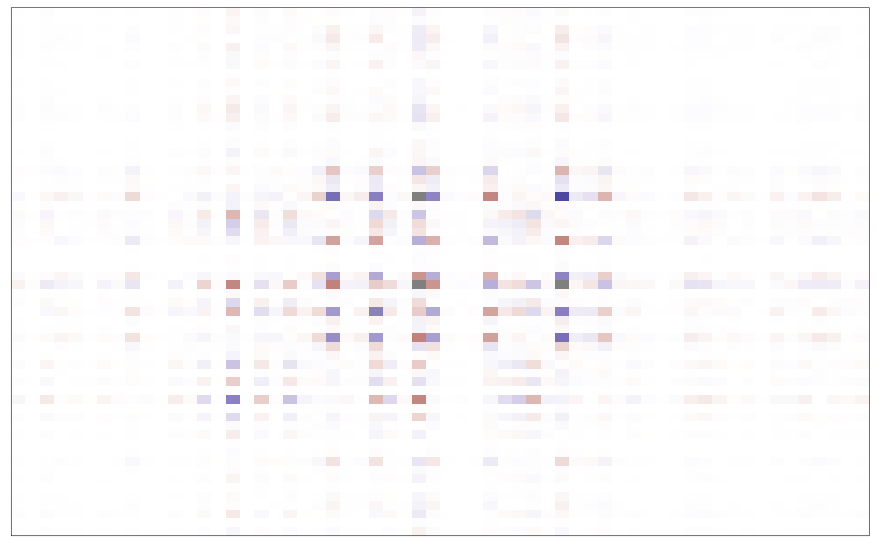}
\end{tabular}
\end{center}
 \caption{Comparison of heatmaps for the estimated (right) and true (left) partially shared loading covariance (A) and common loading covariance (B). Results shown are based on a single dataset generated using the 80\% sparsity setting of the Scenario 1 simulation, where there are structural differences between the common and partially shared factors. Blue indicates negative values, and red indicates positive values.}
\label{fig:heatmaps}
\end{figure}

Tetris was able to differentiate between the common and partially shared signal by correctly identifying high loadings on the appropriate set of features in both types of signal. As an illustration, this can be visually seen in the heatmaps of the estimated common and partially shared loading matrix covariances, as compared to the heatmaps of the true common and partially shared loading matrix covariances (Figure~\ref{fig:heatmaps}). The common loading matrix covariance is the covariance of the loading matrix when subsetted only to the common factors, and analogously so for the partially shared loading matrix covariance. In the interest of space, we are showcasing example results from one simulation in the 80\% sparsity setting, which we consider the most challenging. Although data was simulated ten times under each setting, we are displaying here the run with sixth best performance out of ten (as measured by RV coefficient) in an effort to be representative. Nevertheless, we obtained similar results across the settings and runs. This demonstrates that Tetris can capture and differentiate common and partially shared signals across multiple studies in a range of settings.

\begin{figure}[t]
    \centering
    \includegraphics[width=1\textwidth]{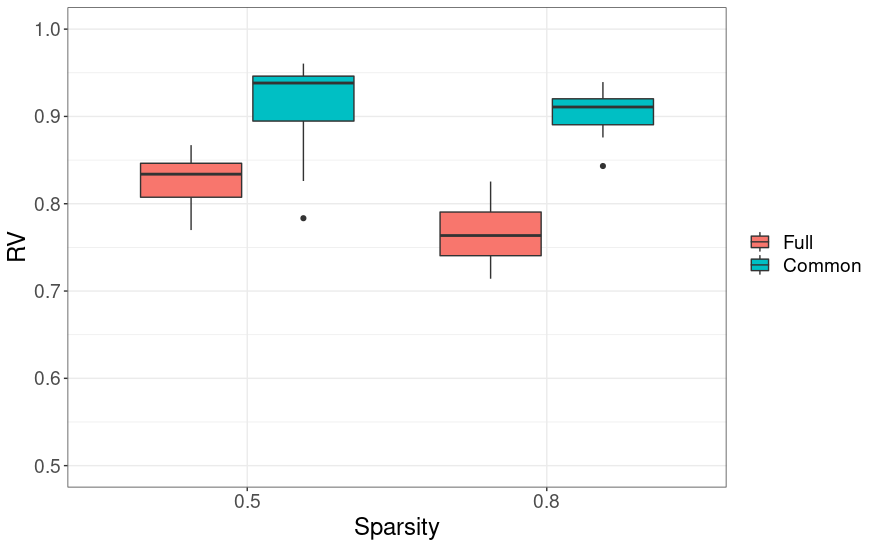}
    \caption{RV coefficients for the full loading matrix covariance and common loading matrix covariance for the Scenario 1 simulation, in both sparsity settings. Colors identify the type of covariance matrix being estimated.}
    \label{fig:RV_heatmap}
\end{figure}

We can also quantify the accuracy of our parameter estimation by looking at the RV coefficient for the full loading matrix covariance (Figure~\ref{fig:RV_heatmap}). The full loading matrix covariance can be interpreted as the loading matrix covariance corresponding to a hypothetical study containing all the factors found across all studies. Intuitively, this quantity summarizes all relationships among variables, across the studies. In this simulation scenario, Tetris's ability to distinguish common and partially shared signal is of particular interest due to the structural differences between the two. Hence, we evaluate this by additionally examining the common loading matrix covariance, which is the covariance when the loading matrix is restricted to the common factors. Analogously to the full loading matrix covariance, this quantity can be thought of as summarizing the structure common to all the studies. For both the full and common loading matrix covariance, we see that the RV coefficients are stable and high over both sparsity settings, confirming that Tetris estimated these parameters well. 

\begin{figure}[t]
    \centering
    \includegraphics[width=1\textwidth]{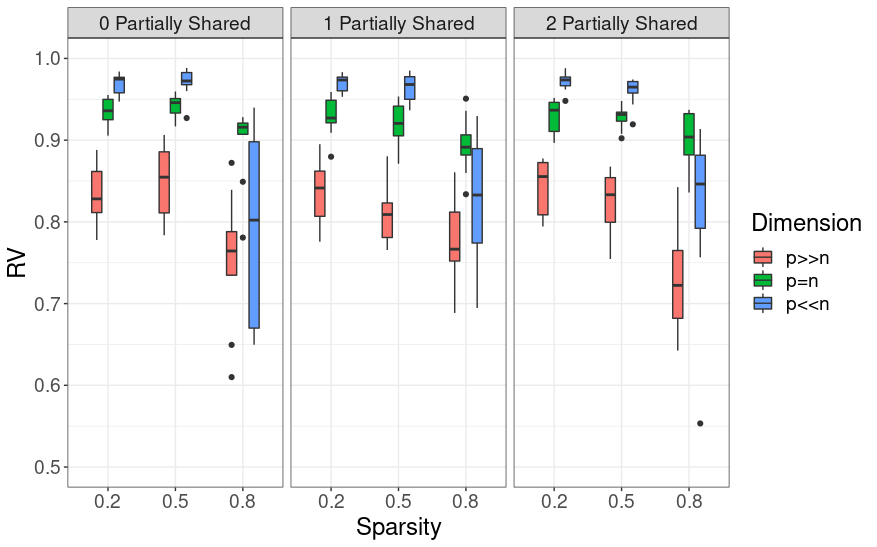}
    \caption{RV coefficients for the full loading matrix covariance across varying sparsity, data dimension, and ground truth number of partially shared factors.}
    \label{fig:full_loading}
\end{figure}

Next, to understand how robust Tetris is under varied settings and without such clear structural differences across factors, we examine the results from the Scenario 2 simulation. The RV coefficients for the full loading matrix covariance are shown (Figure~\ref{fig:full_loading}), demonstrating Tetris's ability to estimate the loading matrix covariance under a wide range of settings. In general, the RV coefficient improves with decreasing sparsity of the loading matrix, as well as with decreasing dimensionality of the data (i.e. $p\ll n$ data results in better estimates than $p\gg n$ data). However, even in challenging scenarios, Tetris remains robust and continues to have reasonable RV coefficients. It is also interesting to note that the RV coefficient does not appear to be affected by the number of partially shared factors, which suggests that Tetris remains effective even when there are more complicated relationships among the studies. 

Thus far, we have shown that Tetris can accurately estimate the loading matrix, which can be thought of as a multi-study parameter. However, we can also demonstrate that Tetris's approach to multi-study estimation has important advantages even when the main quantities of interest are study-specific. In particular, we examine Tetris's ability to recover study-specific signal, by leveraging the common, individual, and applicable partially shared loadings for each study to reconstruct their covariance matrices. We compute the RV coefficients between the reconstructed covariance matrices for each study with the true data-generating covariance matrices in order to assess the accuracy of Tetris's estimates. We also obtain covariance estimates for each study using BMSFA, which only considers common and individual factors, and single-study factor analysis (FA). 

\begin{figure}
\begin{center}
\begin{tabular}{ c c  }
 \textbf{(A)} & \\
 & \includegraphics[width=0.75\textwidth]{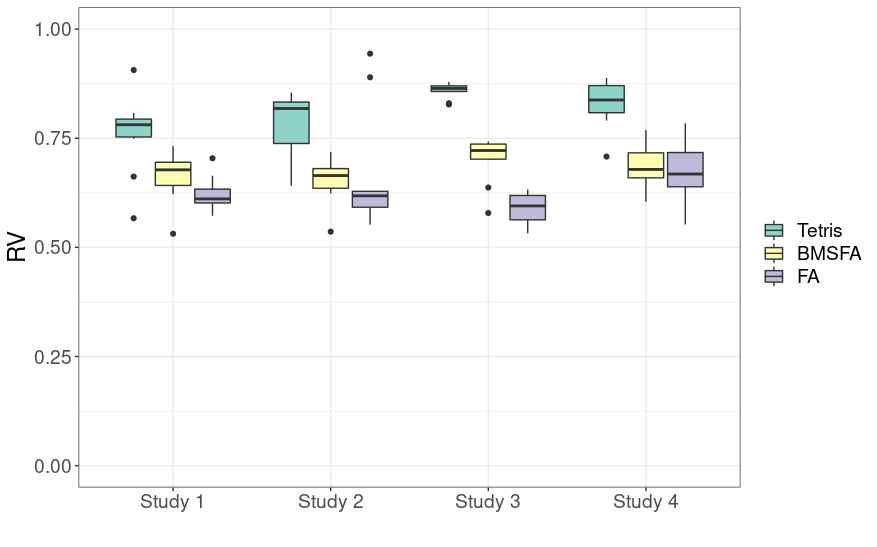}  \\ 
  \textbf{(B)}& \\
  & \includegraphics[width=0.75\textwidth]{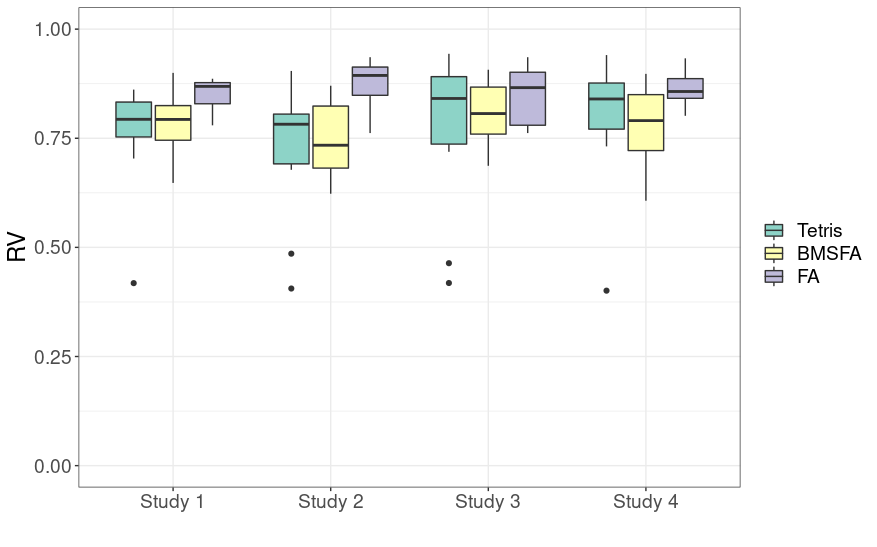} 
\end{tabular}
\end{center}
 \caption{(A) RV coefficients for study-specific covariances in the most challenging scenario ($p\gg n$, 80\% sparsity data with 2 partially shared factors), comparing Tetris to BMSFA and FA. (B) RV coefficients for study-specific covariances in the $p\ll n$, 80\% sparsity setting with 2 partially shared factors. The $p\ll n$, 80\% sparsity scenario is the only setting in which Tetris did not outperform both of the other two methods.}
\label{fig:cov_recovery}
\end{figure}

Tetris recovers these study-specific covariance matrices well (Figure~\ref{fig:cov_recovery}), with the median RV coefficient for each study remaining above 0.75 in each of the 27 parameter combinations tested. In many cases, the median RV coefficient is much higher than 0.75. As we have seen before with the loading matrix covariance, the RV coefficients for Tetris's estimates improve with decreasing loading matrix sparsity and decreasing data dimensionality. In general, Tetris substantially outperforms BMSFA and FA across the parameter settings examined. The fact that Tetris is able to outperform single-study FA shows that joint analysis of multiple studies can improve inference of studies on the study-specific level. The fact that Tetris outperforms BMSFA further shows that partially shared signal can be critical to more accurate estimation. These findings both support the premise of our multi-study, shared factor approach, and suggest that Tetris has utility in covariance estimation. 

The only exception occured in the $p\ll n$, 80\% sparsity setting, where Tetris was comparable to BMSFA and FA in two cases (0 partially factors, and 1 partially shared factor respectively) and performed worse than FA in one case (2 partially shared factors). This is not entirely unexpected, since Tetris is developed especially for the challenging $p\gg n$ setting. We further note that we provided FA with the ground truth number of factors, whereas Tetris and BMSFA both estimated this number on their own. Nevertheless, this observation suggests that Tetris may not always be worthwhile for sparse low-dimensional data, particularly if the number of factors is known a priori. In such cases, more standard methods might be preferable. However, Tetris still yields a high RV coefficient for this scenario, and outperforms both methods in the vast majority of settings, including our setting of greatest interest, sparse $p\gg n$ data. 

\begin{figure}[t]
    \centering
    \includegraphics[width=0.9\textwidth]{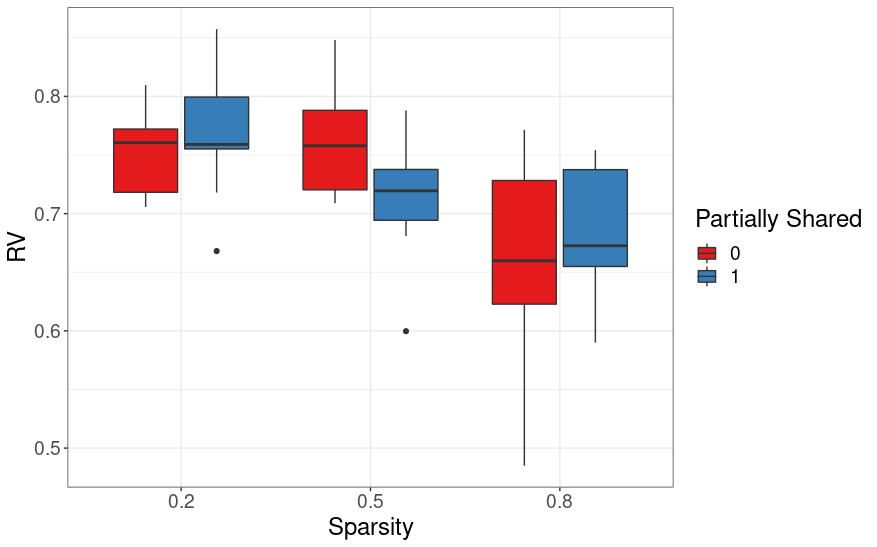}
    \caption{RV coefficients for the full loading covariance for the Scenario 3 simulation, which has 16 studies, across a range of sparsities (0.20, 0.50, and 0.80) and number of partially shared factors (0 or 1), in the $p\gg n$ setting.}
    \label{fig:16studies_loading}
\end{figure}

The findings we have discussed so far focus on the case with four studies, but in practice it is also interesting to consider a setting where a much larger number of studies is available. In the Scenario 3 simulation, we examine 16 studies in the $p\gg n$ setting and assess the accuracy of both the full loading matrix covariance and the study-specific covariances. As with the Scenario 2 simulation, we find that the full loading matrix is estimated well across all sparsity settings (Figure~\ref{fig:16studies_loading}), with performance again improving as sparsity decreases. Even though we have quadrupled the number of studies here, the RV coefficients remain similar in magnitude as when we considered only 4 studies. 

\begin{figure}
\begin{center}
\begin{tabular}{ c c  }
 \textbf{(A)} & \\
 & \includegraphics[width=0.75\textwidth]{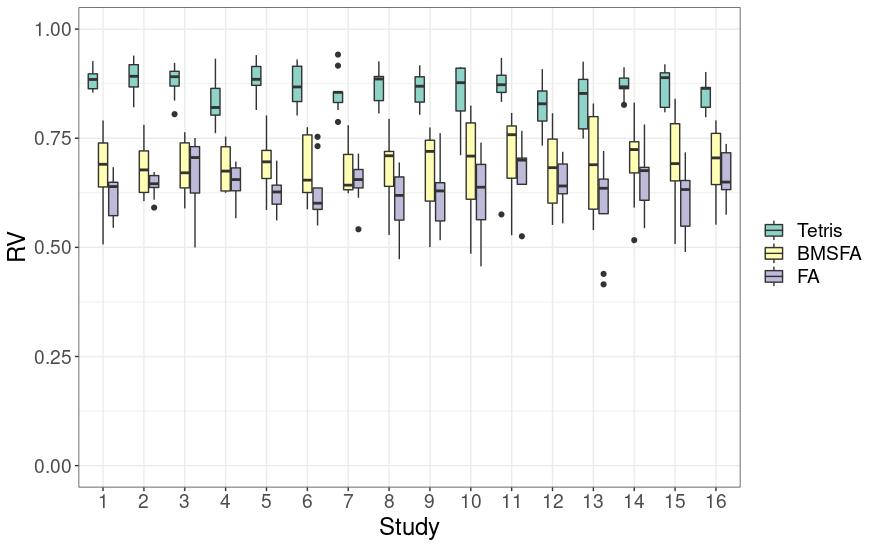}  \\ 
  \textbf{(B)}& \\
  & \includegraphics[width=0.75\textwidth]{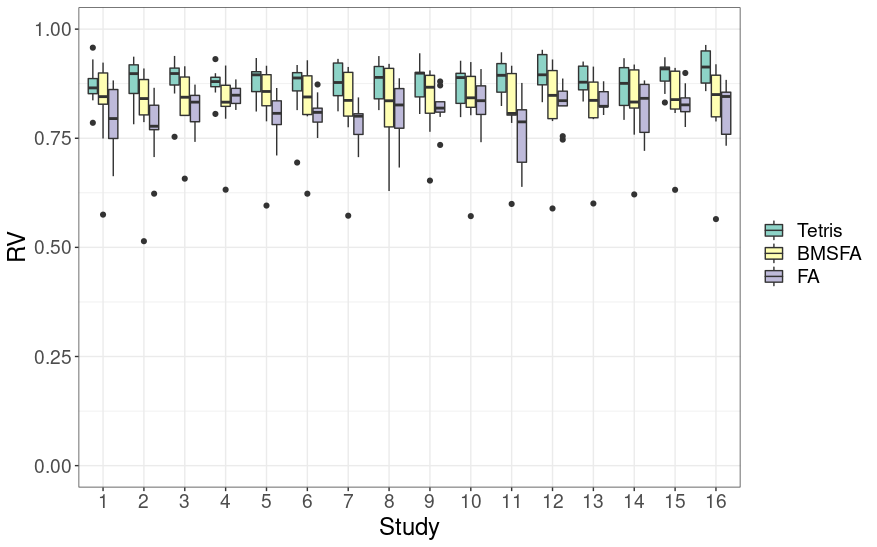} 
\end{tabular}
\end{center}

 \caption{(A) RV coefficients for study-specific covariances in the most challenging case with 16 studies ($p\gg n$, 80\% sparsity data with 1 partially shared factors), comparing Tetris to BMSFA and FA. (B) RV coefficients for study-specific covariances in the easiest case with 16 studies ($p\gg n$, 20\% sparsity data with 0 partially shared factors).}
 \label{fig:16studies_cov}
\end{figure}

We additionally find strong performance for the study-specific covariances (Figure~\ref{fig:16studies_cov}), with Tetris achieving better results than both BMSFA and FA across the board. The advantage from using Tetris is larger in the setting with a partially shared factor (shared by 8 out of the 16 studies) than in the setting with no partially shared factors, as might be expected. Nevertheless, Tetris still represents an improvement in all settings tested, which were restricted to the $p\gg n$ context in this simulation scenario. As an example, we display the results for both the easiest case, i.e. 20\% sparsity, no partially shared factors, and the most challenging case, i.e. 80\% sparsity, one partially shared factor (Figure~\ref{fig:16studies_cov}). These results demonstrate that even with a large number of studies, Tetris retains the ability to accurately estimate the full loading matrix covariance, as well as study-specific covariance matrices. 

\section{Application to RNA-seq Data}

\subsection{Data}

We applied Tetris to RNA-sequencing gene expression data from the immortalized lymphoblastoid cell lines of women at high risk for breast cancer, reported by \cite{pouliot2017transcriptional}. Their experiment produced a total of 121 samples: 37 from subjects with a germline BRCA1 mutation, 50 from subjects with a germline BRCA2 mutation, and 34 from subjects with a strong family history of breast cancer but no BRCA1 or BRCA2 mutation. We refer to these as BRCA1, BRCA2, and BRCAX samples, respectively. Not all subjects in the study were affected by breast cancer. There were 11 affected out of the BRCA1 samples, 18 affected out of the BRCA2 samples, and 17 affected out of the BRCAX samples. 

Our multi-study method can also be used effectively as a multi-condition method. We considered two main ways to define a ``condition" in the context of these data. Firstly, we defined BRCA1, BRCA2, and BRCAX carrier status as the three conditions. Secondly, we defined six conditions by further stratifying each of these mutational categories (BRCA1, BRCA2, and BRCAX) into those who are affected (cases) and unaffected (controls). We refer to these as the unstratified and stratified analyses, respectively. In both, we preprocessed the data in the same way. Namely, we computed transcripts per million (TPM) from raw counts. 
To control computational time, we focus on the 200 transcripts with the highest total TPM summed counts across all samples in the dataset. This is a relatively narrow set of genes, which only captures processes with very high transcriptional levels. These are cell lines derived from lympocytes, so in addition to general cell maintenance and proliferation (e.g. ribosomal proteins), it includes production of immunoglobuline and major glycoproteins found on the surface of immune cells. This is a rather narrow view of the transcriptome, but serves the purpose of illustrating our methodology in a more manageable setting.

We then transformed the data as $\log(TPM+1)$ to approximate normality, an assumption that is met reasonably well in this subset of highly expressed transcripts.
We did not perform any additional preprocessing of the data, such as batch adjustment \citep{Zhang2020.01.13.904730}.
The data may thus retain some batch structure. Batch effects that are confounded with the conditions investigated will likely result in condition-specific factors.
Finally, when applying Tetris, we used the same hyperparameters as described for the simulations.

\subsection{Unstratified analysis}

\begin{figure}
\centering
\includegraphics[width=.95\textwidth]{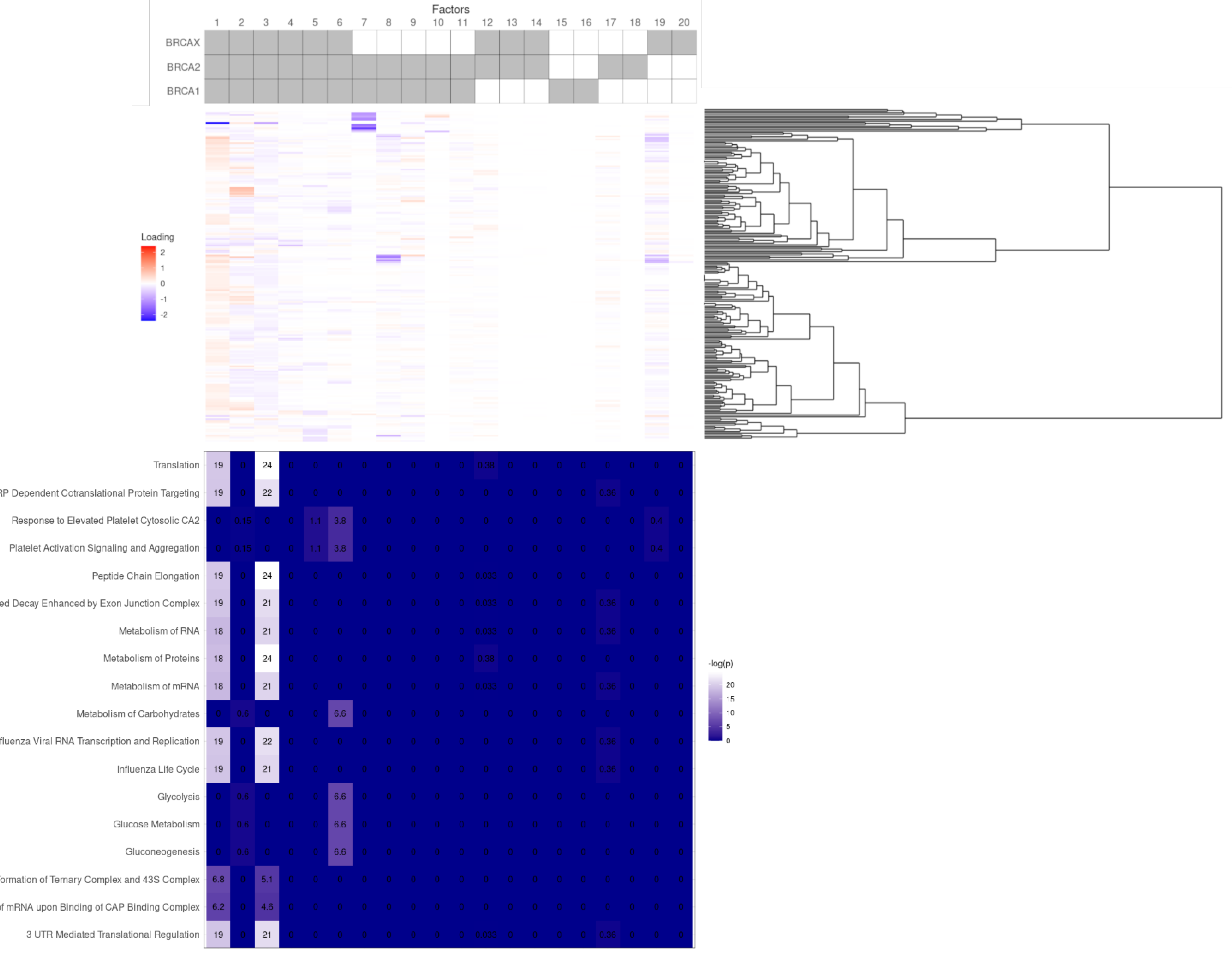}]
\caption{Visual summary of sharing patterns (top), factor loadings (middle) and pathway analysis (bottom) for the unstratified analysis. Each column corresponds to the same factors through the three panels. Transcripts (rows) in the heatmap of loadings are clustered by their raw counts across all samples. Pathways are clustered by gene overlap. Enrichment $p$-values have been corrected with the Benjamini-Hochberg method.}
\label{fig:factor_heatmap}
\end{figure}

In the unstratified analysis (Figure~\ref{fig:factor_heatmap}), the point estimate for the factor indicator matrix had 20 factors in total: two unique factors for each condition, six common factors, five factors shared only by BRCA1 and BRCA2, and three factors shared only by BRCA2 and BRCAX. The larger number of factors shared by BRCA1 and BRCA2 is consistent with what we might expect, since we might anticipate similarities among women in those risk groups. It is interesting to note that there were factors common to all three groups, and factors shared by BRCA2 and BRCAX, but no factors shared by just BRCA1 and BRCAX. Within the credible ball for this point estimate, only approximately $7\%$ of the factor indicator matrices contain factors shared by just BRCA1 and BRCAX. This suggests some similarities between BRCA2 and BRCAX that are not present in BRCA1. This is aligned with epidemiological evidence indicating that BRCA2 carriers are at high risk of other cancers, such as pancreatic cancer, which are not strongly associated with BRCA1, and also have independent, though largely undiscovered, sources of familial risk \citep{doi:10.1043/1543-2165-133.3.365}.

The factor loadings are visualized in Figure~\ref{fig:factor_heatmap}. Several factors, such as the first three common factors, appear to have relatively high loadings across a broad range of transcripts, while others appear to have more concentrated signal in a smaller set of transcripts. For example, the 7th factor, which is shared by BRCA1 and BRCA2, has very large loadings on a set of ten transcripts. Most of these are related to the production of Immunoglobulin (IG). Two specifically (IGHA1 and IGLC2) have been previously identified as biomarkers for improved prognosis in triple-negative breast cancer \citep{hsu2019six}, a subtype of breast cancer most strongly associated with these inherited mutations. Our finding suggests that signal involving these genes may be specific to those with germline BRCA1 and BRCA2 mutations. 

To further interpret these factors, we tested for gene set enrichment \citep{subramanian2005gene} of the factor loadings in the recovered loading matrix. This analysis checks if gene sets representing known biological classes and pathways are enriched in any given factor's loadings above what would be expected at random. We used gene sets from \texttt{reactome.org}, and assessed enrichment with the \texttt{R} package \texttt{RTopper} \citep{marchionni2013rtopper}. 
We found three factors with significant ($p < 0.05$ after Benjamini-Hochberg correction) gene set enrichment. These pathways are shown in Figure~\ref{fig:factor_heatmap}. All of these are common factors. It is notable that two of them are significantly enriched in the same set of pathways, which largely relate to transcriptional and translational processes. They two factors are not highly correlated with one another ($r=0.22$), suggesting that the factor decomposition is able to detect a higher level of resolution compared to the gene set enrichment alone in this case. These processes are known to play an important and broad role in all cellular activities, so it is consistent with prior knowledge that these are enriched in common factors, i.e. relevant to all three groups. The third factor is significantly enriched in pathways related to carbohydrate metabolism and glucose processing. This is particularly interesting, because there have been many investigations into the relationships between carbohydrates, glycemic index, and glycemic load with breast cancer \citep{schlesinger2017carbohydrates,mullie2016relation}. Hence, this result suggests that there may be indeed be an important relationship with such processes and cancer predisposition present in all three groups.

\begin{figure}
\begin{center}
\begin{tabular}{ c c }
    (A) Unstratified (BMSFA) & \\
    \includegraphics[width=0.9\textwidth]{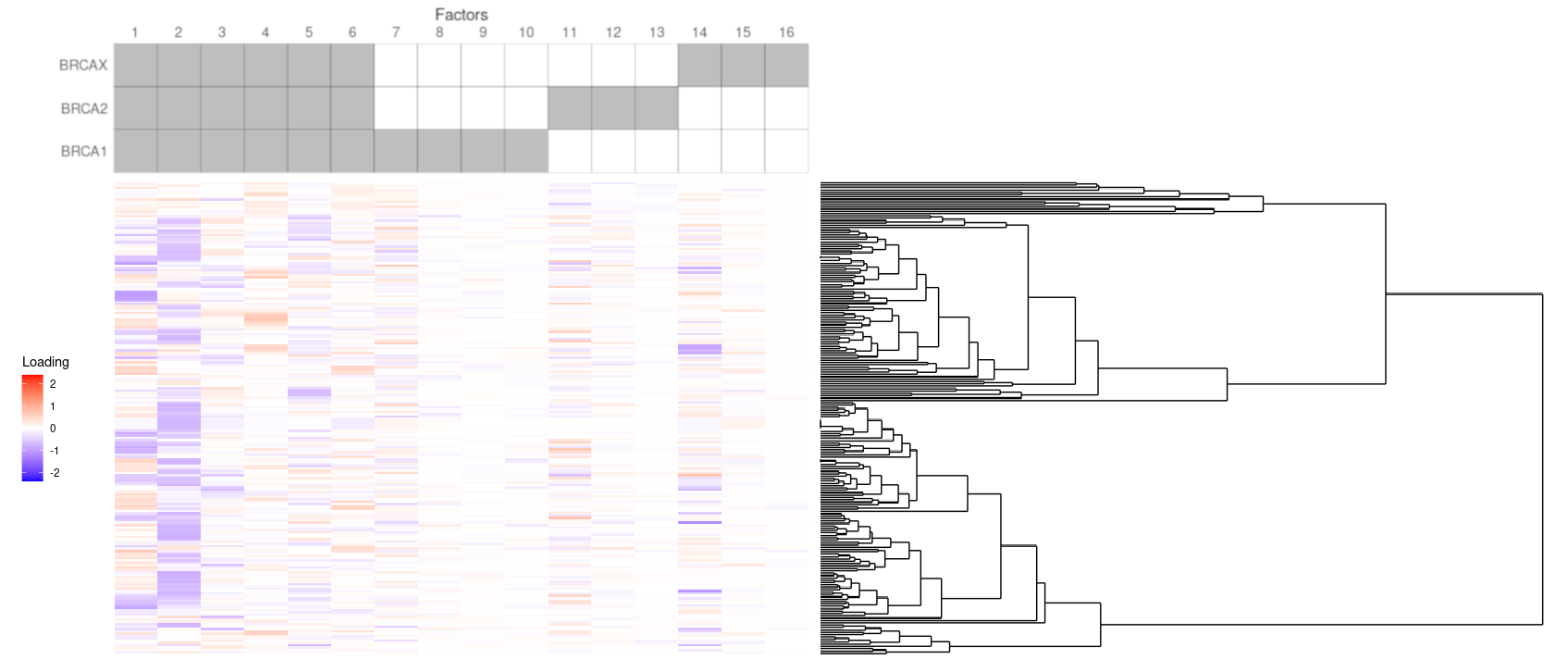} & \\
    (B) Stratified (BMSFA) & \\
    \includegraphics[width=0.9\textwidth]{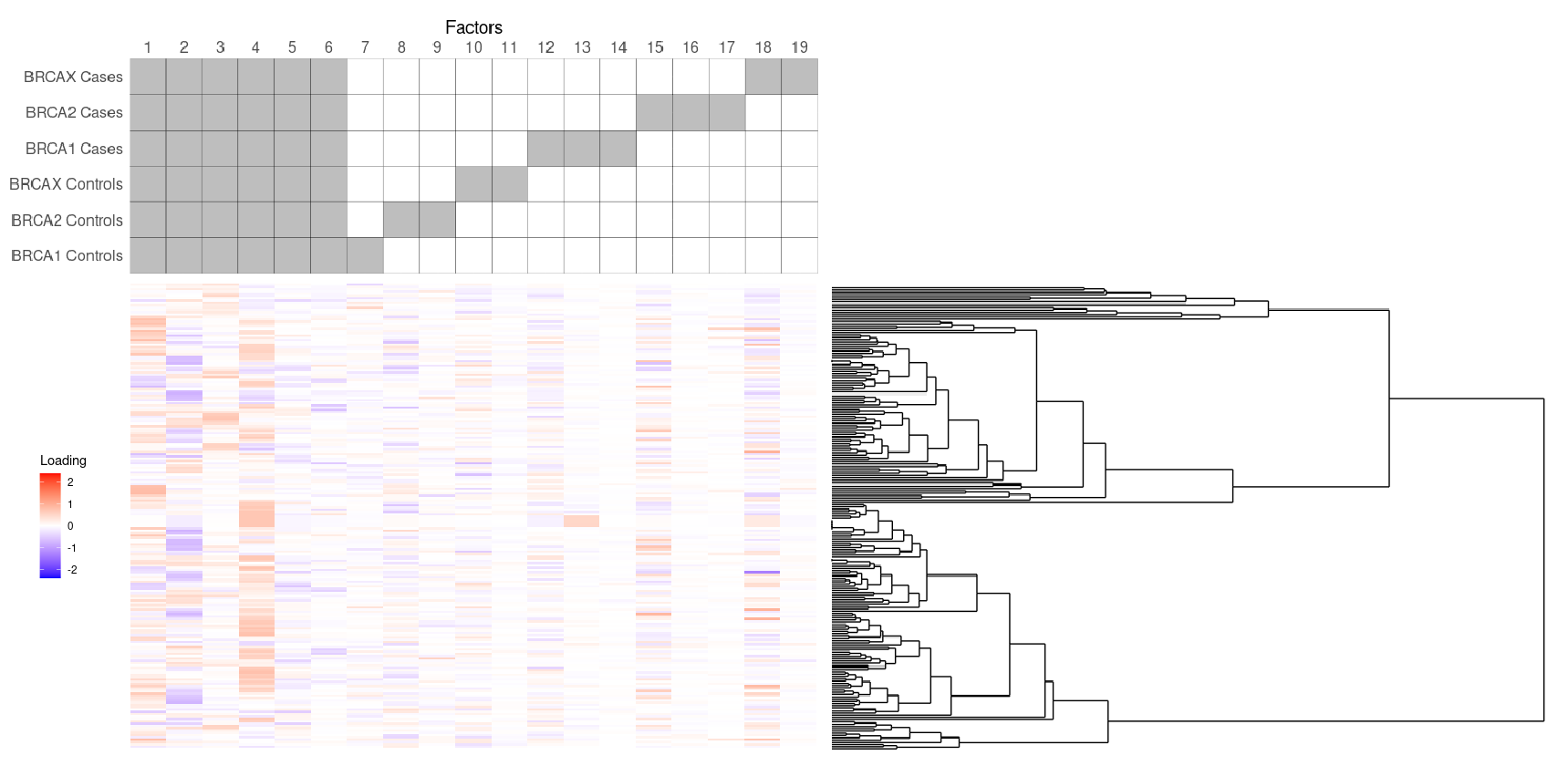} & \\
\end{tabular}
\end{center}
\caption{Heatmap of factor loadings with BMSFA applied to (A) the unstratified  and (B) the stratified analysis.}
\label{fig:factors_bmsfa}
\end{figure}

When repeating this analysis with BMSFA, which only finds common and study-specific factors, we find six common factors, four specific to BRCA1, three specific to BRCA2, and three specific to BRCAX. Many of these factors appear to have broad signals across the transcripts (Figure~\ref{fig:factors_bmsfa}), and some of these factors have fairly strong correlations with the factors found by Tetris. Three of the BMSFA's common factors (1, 4, and 6) are correlated with three of Tetris's common factors (2, 6, and 4 respectively) with $|r| > 0.7$. This includes the factor in Tetris that was significantly enriched in pathways related to carbohydrate metabolism. This suggests a fair amount of overlap in common signal between the two methods. There is also strong correlation ($r = 0.81$) between the 11th factor from BMSFA and the 17th factor from Tetris, which are both factors specific to BRCA2. This is another instance of concordance between the two methods. 

There are also some ways in which Tetris appears to capture similar but more nuanced information than BMSFA. For example, the 12th factor from Tetris, which is shared by both BRCA2 and BRCAX, is correlated ($r = -0.75$) with the 14th factor from BMSFA, which is specific to BRCAX. While it seems that both methods agree this signal belongs to BRCAX, Tetris is able to identify that this signal is also shared by BRCA2. Similarly, the 11th factor from Tetris, which is shared by BRCA1 and BRCA2, is somewhat weakly correlated ($r = -0.57$ and $-0.54$ respectively) with BMSFA's 7th (specific to BRCA1) and 12th (specific to BRCA2) factors. Just examining BMSFA's factor loadings may have led to the hypothesis that these two factors could be the same signal shared by BRCA1 and BRCA2, but Tetris's results lead to this conclusion in a more principled way. 

There are also differences between the results from Tetris and BMSFA. For both of these methods, there are multiple factors that do not appear correlated with any factors in the other. For example, the 7th factor from Tetris, which had particularly high loadings on some transcripts, does not seem to have been present in BMSFA's results.


\subsection{Stratified analysis}

\begin{figure}
    \centering
    \includegraphics[width=0.95\textwidth]{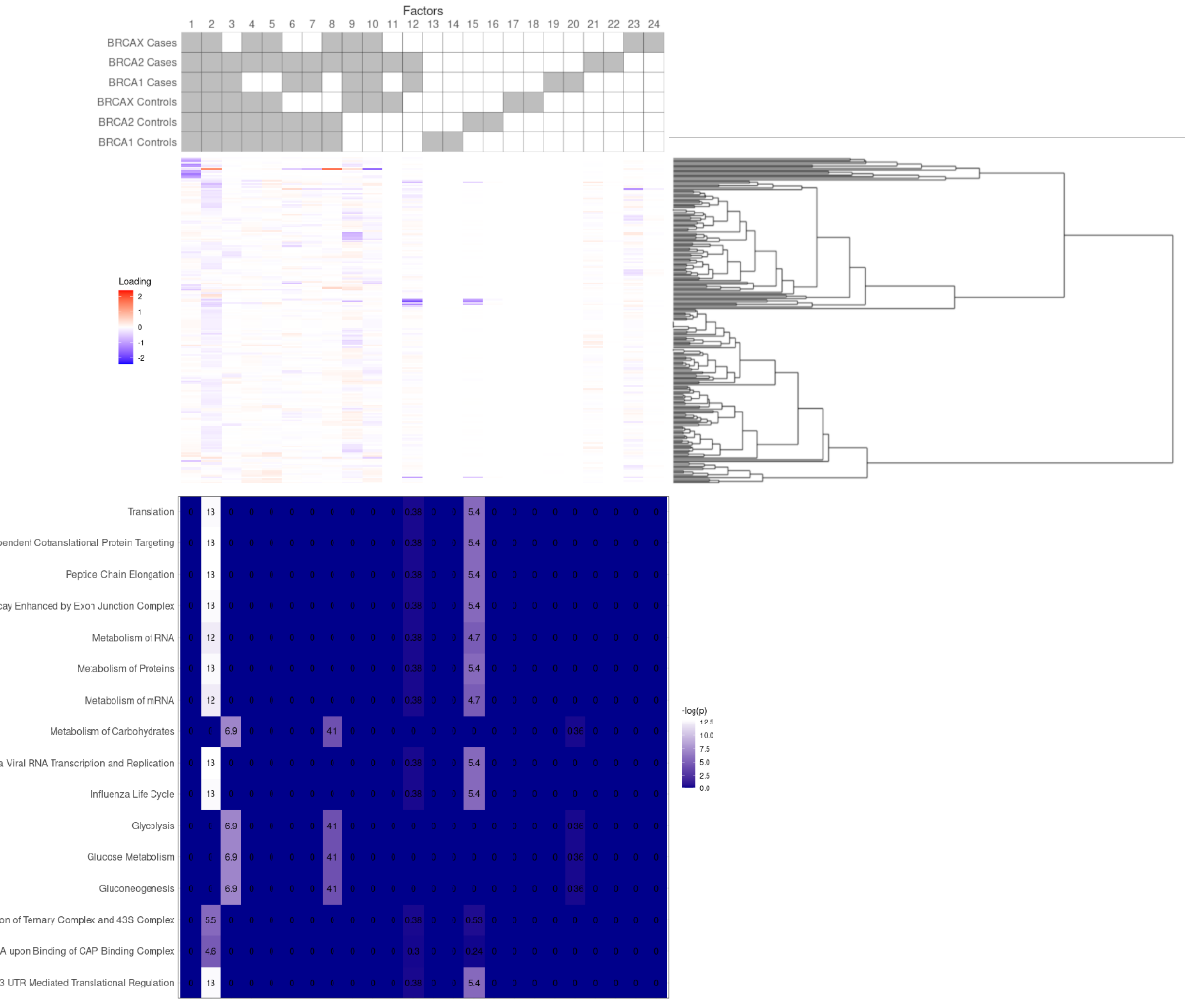}
    \caption{Visual summary of sharing patterns (top), factor loadings (middle) and pathway analysis (bottom) for the stratified analysis. Each column corresponds to the same factors through the three panels. Transcripts (rows) in the heatmap of loadings are clustered by their raw counts across all samples. Pathways are clustered by gene overlap. Enrichment $p$-values have been corrected with the Benjamini-Hochberg method.}
    \label{fig:factor_heatmap_stratified}
\end{figure}

Next, we examine the results from running Tetris in the stratified setting, which further refines the ideas from the unstratified analysis (Figure~\ref{fig:factor_heatmap_stratified}). The point estimate for the factor indicator matrix had 24 factors in all: two study-specific factors for each condition, two common factors, and ten partially shared factors. Of the partially shared factors, two were shared by all except the BRCA1 cases, two were shared by all cases plus BRCAX controls, two were shared by all BRCA1 and BRCA2 groups, one was shared by all except the BRCAX cases, one was shared by BRCA1 and BRCA2 cases, one was shared by all except BRCAX controls and BRCA1 cases, and one was shared by BRCA2 affected and BRCAX unaffected. As with the unstratified analysis, there were no factors shared only by BRCA1 and BRCAX subgroups. Within the credible ball, only $7\%$ of factor indicator matrices contained such factors. Hence, again, there seems to be strong evidence against signal shared only by these two groups. In addition, it is notable that there were more partially shared factors and fewer common factors here than in the unstratified analysis. This can likely be attributed to the fact that the stratification led to greater resolution in distinguishing partially shared signal from common signal. 

When comparing the factors found in the stratified setting to those found in the unstratified setting, it can be seen that many of the stratified factors are consistent with the unstratified factors. For example, the 2nd stratified factor and 1st unstratified factor were highly correlated ($r = -0.89$), and both are common factors. Similarly, the 6th stratified factor (shared by both BRCA1 subgroups and both BRCA2 subgroups) was correlated ($r = -0.72$) with the 9th unstratified factor (shared by BRCA1 and BRCA2). Both of these illustrate agreement even in the stratified setting's higher resolution. 

There are also several cases in which the stratified factors offer a refinement of the unstratified factors. For example, there is very high correlation ($r = 0.97$) between the 21st stratified factor (specific to BRCA2 cases) and the 17th unstratified factor (specific to BRCA2). In this case, the stratified analysis is demonstrating that this signal particularly is present in the cases within BRCA2 samples, but not the controls, whereas the unstratified analysis might suggest the signal is present in both. 
Another example is the strong correlation between the 8th unstratified factor (shared by BRCA1 and BRCA2) and both the 12th stratified factor (shared by both BRCA1 and BRCA2 cases, $r = 0.99$) and the 15th stratified factor (specific to BRCA2 controls, $r = 0.92$). One of the transcripts with high loadings on these factors corresponds to the gene EEF1A1, which has been previously implicated in epigenetic regulation of breast cancer cells \citep{li2019genome}. Hence, the unstratified analysis suggests that this gene and those whose regulations it controls may play an important role among those with BRCA1 or BRCA2 germline mutations. However, the stratified analysis refines this idea in two ways. First, it suggests that BRCA1 controls are not sharing this signal, since both the 12th and 15th factors do not belong to BRCA1 controls. Second, by identifying these two separate factors, it points to a subtle difference in the way the signal manifests in BRCA1 and BRCA2 cases versus in BRCA2 controls. 

We further interpreted the set of stratified factors by assessing gene set enrichment, implemented as earlier. In this setting, four factors had significant ($p < 0.05$ after Benjamini-Hochberg correction) gene set enrichment, with the pathways shown in Figure~\ref{fig:factor_heatmap_stratified}. Of these factors, one is study-specific to BRCA2 controls, one is a common factor, one is the partially shared factor belonging to all except the BRCAX cases, and one is the partially shared factor belonging to all except the BRCAX cases and BRCA1 controls. 

The common factor was significantly enriched for the same set of pathways related to transcription and translation as two of the common factors in the unstratified analysis. This is expected, since, as we previously noted, there is a strong correlation between this common factor and the first common factor in the unstratified setting. The study-specific factor was significantly enriched for a large subset of these pathways as well. This again supports the idea that transcriptional and translational processes play a broad role across all the groups under consideration.

The two partially shared factors were both significantly enriched for the same set of glucose- and carbohydrate-related processes as the 6th common factor in the unstratified analysis. However, it should be noted that only one of these (the 3rd stratified factor, which is shared by all but BRCAX cases) is correlated with this unstratified factor ($r = 0.85$). Whereas these processes previously appeared to be relevant for all groups in the unstratified setting, the significant enrichment in this 3rd stratified factor refines this idea to suggest that these processes do not play the same role within BRCAX cases. Since the other stratified factor is not especially correlated, this may indicate an entirely different signal from what the unstratified analysis detected, even though the same set of pathways are enriched.

When comparing Tetris's results in the stratified setting to those of BMSFA (Figure~\ref{fig:factors_bmsfa}), there are fewer correspondences than in the unstratified setting. The 9th factor from Tetris (shared by all but BRCA1 controls and BRCA2 controls) is correlated with the 2nd common factor from BMSFA ($r = 0.77$), which can be interpreted as a refinement of BMSFA's finding. Similarly, the 21st factor for Tetris and the 15th factor for BMSFA are strongly correlated ($r = 0.83$), which makes sense since both of these are specific to BRCA2 cases. Otherwise, there aren't any other pairs of factors with correlations over 0.70. This might be because the larger number of conditions results in a larger number of possibilities for partial sharing, and hence we might expect less simple agreement than in the unstratified setting.

\section{Discussion}
We have presented the first multi-study factor analysis method that permits factors to be shared by any subset of studies. Tetris can be applied not only to datasets from multiple studies, but also data from multiple conditions within a study. We tested Tetris on a range of simulations and demonstrated its accuracy in estimating model parameters and distinguishing common, study-specific, and partially shared signal. This decomposition of signal offers a precise way to quantify wholly or partially shared information across studies, avoiding the otherwise common practice of running separate analyses on each study and subjectively integrating the results. Tetris instead offers a more principled form of joint dimension reduction that directly estimates each factor and its membership across studies. Moreover, our use of the Indian Buffet Process prior results in automatic dimension selection, without necessitating ad-hoc post-processing steps. 

We further highlighted how this flexible approach to joint estimation allows Tetris to borrow strength across studies and recover study-specific covariance matrices more accurately than standard single-study factor analysis. Thus, our approach offers an opportunity to leverage multiple studies to improve analysis even for research questions pertaining to a single study. Since Tetris also outperformed BMSFA on study-specific covariance estimation, we have demonstrated that permitting partial sharing of factors can result in substantial improvements over just the consideration of common and study-specific factors. 

Our approach is similarly effective in jointly analyzing multivariate data collected in multiple conditions. 
We applied Tetris to gene expression datasets to jointly identify signal among women with known breast cancer mutations and high familial risk.  The identification of partially shared factors provided a more detailed understanding of how signal is partitioned across the multiple conditions, which would have been lost if only common and condition-specific factors were estimated. In particular, we recovered numerous instances of signal sharing between those with BRCA1 and BRCA2 germline mutations that were not present in those without the mutations, even though there were other factors common to all. When we further stratified these groups into those who are affected and unaffected by breast cancer, the result was an even higher-resolution understanding of whether this signal was associated with anyone having these mutations, or e.g. only after onset of cancer. This analysis recapitulated known biology by identifying signal related to genes previously associated with breast cancer. By specifically implicating these genes in factors shared by a subset of conditions, our analysis then suggests novel hypotheses about whether these genetic mechanisms of action may be particularly relevant for certain subgroups. Hence, we have shown how Tetris can be successfully employed in the unsupervised analysis of complex, multi-study or multi-condition data. This is useful for a wide range of applications beyond genomics, such as epidemiology, nutrition, and sociology. 


One of the key goals of our inference is the factor indicator matrix $\mathcal{A}$. We suggest to consider point estimates based on defining a distance and identifying the point defining the neighborhood with highest density. To do so, we developed a notion of distance for comparing binary matrices of this nature, and drew a connection with the Hungarian Algorithm to compute it efficiently. There are many possible alternative options to summarize the MCMC samples of $\mathcal{A}$ with a single point estimate. For example, the maximum a posteriori value could be selected, where the posterior density $f(\mathcal{A}|\boldsymbol{X})$ for data $\boldsymbol{X}$ is approximated using importance sampling or variational inference. There is also a wider literature of Bayesian model averaging \citep{hoeting1998bayesian} and Bayesian model selection \citep{chipman2001practical} that could be considered and adapted to this problem, including approaches based on Bayes factors \citep{berger1996intrinsic}, Bayesian information criteria \citep{chen2008extended}, and techniques from reversible jump MCMC \citep{lopes2004bayesian}. For our purposes, we found our chosen approach was simple and achieved strong results. We also adapted the idea of credible balls \citep{wade2018bayesian} from Bayesian clustering to express regions of uncertainty for our point estimate. This offers an important summary of uncertainty that is useful not only in our specific context, but in any model considering infinite binary matrices such as those under the Indian Buffet Process. 

Simultaneous factorization of multiple matrices sharing common structure is an area of active research. For example,
\cite{roy2019perturbed} present Perturbed Factor Analysis (PFA), in which multiple groups or studies are modeled as arising from perturbations to a common factor loading matrix. It is useful to contrast their approach to ours. While their method addresses a related question, we could consider PFA, BMSFA, and our Tetris as belonging to a continuum in some sense. PFA might be most useful in contexts where multiple studies are believed to have very strong common structure, without interest in explicitly estimating study-specific factor loadings. For example, this could be the case in an application where the goal is to remove numerous and small batch effects, which are not of direct interest, from common signal. BMSFA might instead be most applicable when both the common and study-specific structure are of particular interest, when the batch effects are strong and highly confounded with study, and when partially shared structure is not likely to exist. An example might be when each study is based in a different subpopulation, and it is important to understand the unique contributions from each subpopulation. Finally, through Tetris, we contribute an approach that is most useful when partially shared structure is plausible, and when explicit estimates of each type of signal are desired. This might be the case if we expect subsets of our studies to share commonalities without being identical. For example, in our gene expression case study, we expect some similarities between BRCA1 and BRCA2 samples. We might then be interested in answering questions about what all groups share, what is unique to each group, and what BRCA1 and BRCA2 samples might share that does not belong to BRCAX samples. Hence, these three approaches (PFA, BMSFA, and Tetris) can be understood as existing along a continuum for complementary but not identical contexts.


In summary, Tetris demonstrates that it is viable to perform flexible joint analyses of multiple high-dimensional studies, identifying common, study-specific, and partially shared structure. Multi-study unsupervised analyses are underutilized, despite the fact that the multi-study setting offers a unique foundation for validation of the signal identified. We hope that this work will support a broader application of multi-study methods in unsupervised learning applications.

\section*{Acknowledgements}
Research supported by the U.S.A.'s National Science Foundation grant NSF-DMS1810829 (LT and GP), the U.S.A.'s National Institutes of Health grant NCI-5P30CA006516-54 (GP), the National Library Of Medicine of the National Institutes of Health under Award Number T32LM012411 (ING), and the National Science Foundation Graduate Research Fellowship under Grant No. DGE1745303 (ING). Any opinions, findings, and conclusions or recommendations expressed in this material are those of the author(s) and do not necessarily reflect the views of the National Science Foundation. The content is solely the responsibility of the authors and does not necessarily represent the official views of the National Institutes of Health.

\bibliographystyle{apalike}  
\bibliography{main}

\end{document}